\documentstyle[amsfonts,12pt]{article}

\begin{document}

\title{Off-Shell Formulation of Simple Supersymmetric Yang-Mills}
\author{Khaled Abdel-Khalek \thanks{%
khaled@le.infn.it} \thanks{%
Address after 15 Feberuary 2000: Feza G\"ursey Institute, P.O. Box 6, 81220 
\c{C}engelk\"oy, \'Istanbul, Turkey} \\
Dipartimento di Fisica, Universit\`a di Lecce\\
- Lecce, 73100, Italy -}
\date{February 2000}
\maketitle

\begin{abstract}
An off-shell formulation for 6 and 10 dimensions simple supersymmetric
Yang-Mills theories is presented. While the fermionic fields couple to left
action of \ $S^{3}$ and $S^{7}$ respectively, the auxiliary ones couple to
right action (and vice versa). To close the algebra off-shell, left and
right actions must commute. For 6 dimensions quaternions work fine. The 10
dimensional case needs special care. Pure spinors and soft Lie algebra
(algebra with structure functions instead of structure constants) are
essential. Some tools useful for constructing the superspace are also
derived. We show how to relate our results to the early works of Evans and
Berkovits.
\end{abstract}

\newpage

\section{Introduction}

Day after day, supersymmetry consolidates its position in theoretical
physics. Even if it was introduced more than 25 years ago, there are still
problems with the geometric basis of extended $\left( N>1\right) $
supersymmetry. The situation of the extended superspace is far less
satisfactory than the original N=1 superspace. At the level of the algebra
the on-shell formalism closes up to modulo of the classical equations of
motion. This fact seems odd at the quantum level since the equations of
motion receive loop corrections\footnote{%
Also, the supersymmetry transformations receive corrections and one should
test the closure of \ the algebra order by order in perturbation theory.}.
The superspace introduces an elegant supermanifold with different enlarged
superconnections, where some are truly integrable in the sense of having
zero supercurvature. In principle, the extended superspace should be a very
powerful tool for quantum calculations.

Before starting, we feel obliged to mention something about the history of
the following conjecture: Ring Division Algebras ${\Bbb K}\equiv $\{ real $%
{\Bbb R},\;$complex ${\Bbb C},\;$quaternions ${\Bbb H},\;$octonions ${\Bbb O}
$ \} are relevant to simple supersymmetric Yang-Mills. The first hint, as
mentioned by Schwarz \cite{r1} comes from the number of propagating Bose and
Fermi degrees of freedom which is one for $d=3$, two for $d=4$, four for $%
d=6 $ and eight for $d=10$ suggesting a correspondence with real ${\Bbb R}$,
complex ${\Bbb C}$, quaternions ${\Bbb H}$ and octonions ${\Bbb O}$. Kugo
and Townsend \cite{r2} investigated in detail the relationship between $%
{\Bbb K}$ and the irreducible spinorial representation of the Lorentz group
in $d=3,4,6,10$, building upon the following chain of isomorphisms 
\begin{eqnarray*}
so\left( 2,1\right) &\Longleftrightarrow &sl(2,{\Bbb R}) \\
so(3,1) &\Longleftrightarrow &sl(2,{\Bbb C}) \\
so(5,1) &\Longleftrightarrow &sl(2,{\Bbb H}).
\end{eqnarray*}
They conjectured that $so(9,1)\Longleftrightarrow sl(2,{\Bbb O})$, the
correct relation turned out to be 
\[
so(9,1)\Longleftrightarrow sl(2,{\Bbb O})\oplus G_{2} 
\]
as has been shown by Chung and Sudbery \cite{r3}, i.e. the dimension of $%
Sl\left( 2,{\Bbb O}\right) $ is 31. Also in \cite{r2}, a quaternionic
treatment of the $d=6$ case is presented. Later, Evans made a systematic
investigation of the relationship between SSYM and ring division algebra in
a couple of papers. In the first \cite{r4}, he simplified the construction
of SSYM by proving a very important identity between gamma matrices by using
the intrinsic triality of ring division algebra instead of the ``tour de
force'' used originally by Brink, Scherck and Schwarz \cite{r0} via Fierz
identities generalized to $d>4$ dimensions. Then, in the second paper \cite
{r6}, Evans made the connection even clearer by showing how the auxiliary
fields are really related to ring division algebras. For $d=3,4,6,10$ we
need $k=0,1,3,7$ auxiliary fields respectively. An alternative approach for
the octonionic case was introduced by Berkovits \cite{r7} who invented a
larger supersymmetric transformation called generalized supersymmetry in 
\cite{r8}. There has also been a twistor attempt by Bengtsson and Cederwall 
\cite{r9}. For more references about the octonionic case and ten dimensional
physics one may consult references in \cite{r10} and its extension to
p-branes by Belecowe and Duff \cite{r11}. The early work of Nilsson may be
relevant \cite{r12}\cite{r13} too.

As a first step towards an extended superspace, we address the point of the
algebraic auxiliary fields for simple N=1 supersymmetric Yang-Mills (SSYM)
definable only in $d=3,4,6$ and $10$ dimensions \cite{r0}. The important
point is: {\em While the physical fields couple to ring division left action
the auxiliary ones couple to right action (or vice versa)}. To admit a
closed off-shell supersymmetric algebra, left and right action must commute
i.e. we should have a parallelizable associative algebra. For d = 6,
quaternions work fine but for d = 10, the only associative seven dimensional
algebra that is known is the soft seven sphere. We shall show below how this
works. In this work, we use the same symbols $\left( \mbox{left action}%
\equiv {\Bbb E}_{j}\mbox{,
right action}\equiv 1|{\Bbb E}_{j}\right) $ for either complex, quaternionic
or octonionic numbers and each case should be distinguished by the range of
the indices $j$ which run from $1$ to $\left( 1,3,7\right) $ for complex,
quaternions and octonions respectively.

In the second section, we review the relation between hypercomplex structure
and Clifford algebra. The auxliary fields problem in 6 dimensions is
presented in the third section. While section four is devoted to the ten
dimensional case. The last section contains some superspace hints.

\section{Hypercomplex Structure and Pure Spinors}

Everything starts from Clifford algebra, so let's review quickly the
connection between hypercomplex structures and our gamma matrices in $%
d=3,4,6,10$. The solution is encoded completely in our $\Gamma _{M}$.
Algebraically, we can construct a Clifford algebra directly from complex,
quaternions and octonions over $1,3,7$ Euclidean space which can be extended
easily to a representation of the minimal irreducible spinorial subspaces in 
$d=3,4,6$ and $10$ Minkowskian space-time. Consider the set of matrices $%
\left\{ {\Bbb E}_{j}\right\} $ for the following three different cases\cite
{r14}\cite{r15}:

\begin{itemize}
\item[$\bullet $]  the canonical complex structure over ${\Bbb R}^{2}$ is
just $2\times 2$ matrix ${\Bbb E}_{1}$%
\begin{equation}
e_{1}\Longleftrightarrow {\Bbb E}_{1}=\left( \delta _{0\mu }\delta _{1\nu
}-\delta _{0\nu }\delta _{1\mu }\right) ;\quad \left( {\Bbb E}_{1}\right) =-%
{\bf 1}_{\mu }\quad ;\quad {\footnotesize \mu =0,1\quad ,}
\end{equation}
by ${\bf 1}_{\mu }$ we always mean an ($\mu \times \mu $) unit matrix.

\item[$\bullet \bullet $]  the canonical left quaternionic structures over $%
{\Bbb R}^{4}$ are 
\begin{equation}
({\Bbb E}_{j})_{\mu \nu }=(\delta _{0\mu }\delta _{j\nu }-\delta _{0\nu
}\delta _{j\mu }-\epsilon _{j\mu \nu })\quad \quad {\footnotesize \mu ,\nu
=0..3;\;j,k,h=1..3\quad ,}
\end{equation}
and 
\begin{equation}
{\Bbb E}_{j}{\Bbb E}_{k}=(-\delta _{jk}{\bf 1}_{\mu }+\epsilon _{jkh}{\Bbb E}%
_{h})\qquad ,  \label{q2}
\end{equation}
where $\epsilon _{jkh}$ is the standard Levi-Civita symbol. Using Rotelli's
notation for right action\cite{r16}, the canonical right quaternionic
structures are, 
\begin{equation}
({1|}{\Bbb E}_{j})_{\mu \nu }=(\delta _{0\mu }\delta _{j\nu }-\delta _{0\nu
}\delta _{j\mu }+\epsilon _{j\mu \nu })\quad ,
\end{equation}
and 
\begin{equation}
1|{\Bbb E}_{j}~1|{\Bbb E}_{k}=(-\delta _{jk}{\bf 1}_{\mu }-\epsilon _{jkh}1|%
{\Bbb E}_{h})\qquad .  \label{q4}
\end{equation}
Let's put these quaternionic structures into a form that can be recognized
by physicsits 
\[
\left( {\Bbb E}_{j}\right) _{\mu \nu }=-\left( 1|{\Bbb E}_{j}\right) _{\mu
\nu }=-\epsilon _{j\mu \nu }\;\;\;if\;\;\;\;{\footnotesize \mu ,\nu =1,2,3}%
.\quad \left( {\Bbb E}_{j}\right) _{00}=\left( 1|{\Bbb E}_{j}\right)
_{00}=0\quad .
\]
\begin{equation}
\left( {\Bbb E}_{j}\right) _{0\nu }=\left( 1|{\Bbb E}_{j}\right) _{0\nu
}=-\delta _{j\nu },\qquad \left( {\Bbb E}_{j}\right) _{\mu 0}=\left( 1|{\Bbb %
E}_{j}\right) _{\mu 0}=\delta _{j\mu }\quad ,
\end{equation}
such mathematical quaternionic structures are well known in physics as the
't~Hooft eta symbols\cite{r17}. We can check that 
\[
\left\{ {\Bbb E}_{i,}{\Bbb E}_{j}\right\} =\left\{ 1|{\Bbb E}_{i},1|{\Bbb E}%
_{i}\right\} =-2\delta _{ij}{\bf 1}_{4}\quad ,
\]
\begin{eqnarray}
\left[ {\Bbb E}_{j},{\Bbb E}_{k}\right]  &=&\epsilon _{jkh}{\Bbb E}_{h}\quad
\quad \quad ,  \nonumber \\
\left[ 1|{\Bbb E}_{j},1|{\Bbb E}_{k}\right]  &=&-\epsilon _{jkh}1|{\Bbb E}%
_{h}\quad \quad ,
\end{eqnarray}
and the very important formula 
\begin{equation}
\left[ {\Bbb E}_{i},1|{\Bbb E}_{j}\right] =0\quad ,
\end{equation}
i.e. left and right quaternionic actions commute.
\end{itemize}

{\em For octonions}, the story is quite different, as they are
non-associative. But as it is well known, for any Lie algebra the structure
constants are proportional to the constant torsion over the group manifold
whereas the torsion over the seven sphere $S^{7}$ varies from one point to
another \cite{r18}. The only way to solve these problems is to use the $S^{7}
$ as an associative soft Lie algebra\footnote{%
Soft Lie algebra is an algebra with structure functions instead of structure
constants \cite{r18a}.} as had been proposed by Englert, Servin, Troost,
Van~Proeyen and Spindel \cite{r19} which can be derived from octonions (Look
to \cite{r20} for a full algebraic investigation of the soft seven sphere).
For a generic octonionic number, 
\begin{equation}
\varphi =\varphi ^{\mu }e_{\mu }=\varphi _{0}e_{0}+\varphi
_{i}e_{i,}\Longleftrightarrow \left( 
\begin{array}{c}
\varphi _{0} \\ 
\varphi _{1} \\ 
\varphi _{2} \\ 
\varphi _{3} \\ 
\varphi _{4} \\ 
\varphi _{5} \\ 
\varphi _{6} \\ 
\varphi _{7}
\end{array}
\right) ,\quad 
\begin{array}{lll}
{\footnotesize \mu ,\nu =0..7} & {\footnotesize ,} & {\footnotesize %
j,k,h=1..7} \\ 
&  &  \\ 
& \varphi ^{\mu }\in {\Bbb R} & 
\end{array}
\quad {\footnotesize ,}
\end{equation}
such that $e_{0}=1$ and the other seven imaginary units satisfy $%
e_{i}e_{j}=-\delta _{ij}+f_{ijk}e_{k}\Longleftrightarrow
[e_{i},e_{j}]=2f_{ijk}e_{k}$ where $f_{ijk}$ is completely antisymmetric and
equals one for any of the following three-cycles (123), (145), (246), (347),
(176), (257), (365). To construct the soft seven sphere Lie algebra, we just
have to define the direction of action, for left and right action, we have 
\begin{eqnarray}
\delta _{i}\varphi  &=&e_{i}\varphi \quad ,  \nonumber \\
1|\delta _{i}\varphi  &=&\varphi e_{i}\quad ,
\end{eqnarray}
then after simple calculations, we find 
\begin{eqnarray}
\lbrack \delta _{j},\delta _{k}] &=&2f_{jkh}\delta _{h}-2[\delta
_{j},1|\delta _{k}]\;\quad \quad ,  \nonumber \\
\left[ 1|\delta _{j},1|\delta _{k}\right]  &=&-2f_{jkh}1|\delta
_{h}-2[\delta _{j},1|\delta _{k}]\quad ,  \nonumber \\
\{\delta _{j},\delta _{k}\} &=&-2\delta _{jk}\quad \quad \quad \quad \quad
\quad \quad \quad \quad ,  \nonumber \\
\{1|\delta _{j},1|\delta _{k}\} &=&-2\delta _{jk}\qquad \quad \quad \quad
\quad \quad \quad \quad ,  \label{ll4}
\end{eqnarray}
which are isomorphic to the following set$\ \{{\Bbb E}_{j},1|{\Bbb E}_{j}\}$
of $8\times 8$ matrices, 
\begin{equation}
\begin{array}{ccccc}
\delta _{j} & \Longleftrightarrow  & ({\Bbb E}_{j})_{\mu \nu } & = & \delta
_{0\mu }\delta _{j\nu }-\delta _{0\nu }\delta _{j\mu }-f_{j\mu \nu }\quad ,
\\ 
1|\delta _{j} & \Longleftrightarrow  & (1|{\Bbb E}_{j})_{\mu \nu } & = & 
\delta _{0\mu }\delta _{j\nu }-\delta _{0\nu }\delta _{j\mu }+f_{j\mu \nu
}\quad ,
\end{array}
\end{equation}
satisfying the algebra \cite{r14}
\begin{eqnarray}
\lbrack {\Bbb E}_{j},{\Bbb E}_{k}] &=&2f_{jkh}{\Bbb E}_{h}-2[{\Bbb E}_{j},1|%
{\Bbb E}_{k}]\quad \quad \quad ,  \nonumber \\
\left[ 1|{\Bbb E}_{j},1|{\Bbb E}_{k}\right]  &=&-2f_{jkh}1|{\Bbb E}_{h}-2[%
{\Bbb E}_{j},1|{\Bbb E}_{k}]\quad \;,  \nonumber \\
\{{\Bbb E}_{j},{\Bbb E}_{k}\} &=&-2\delta _{jk}\quad \quad \quad \quad \quad
\quad \quad \quad \quad \quad ,  \nonumber \\
\{1|{\Bbb E}_{j},1|{\Bbb E}_{k}\} &=&-2\delta _{jk}\quad \quad \quad \quad
\quad \quad \quad \quad \quad \quad ,  \label{lll4}
\end{eqnarray}
they don't close a Lie algebra but they close a soft Lie algebra defined by 
\[
\lbrack \delta _{j},\delta _{k}]\varphi \equiv 2f_{jkh}^{\left( +\right)
}(\varphi )~e_{h}\varphi \Longleftrightarrow [{\Bbb E}_{j},{\Bbb E}_{k}]%
{\normalsize \varphi }=2f_{jkh}^{\left( +\right) }{\Bbb E}_{h}{\normalsize %
\varphi \quad ,}
\]
\begin{equation}
\lbrack 1|\delta _{j},1|\delta _{k}]\varphi \equiv 2f_{jkh}^{\left( -\right)
}(\varphi )~\varphi e_{h}\Longleftrightarrow \left[ 1|{\Bbb E}_{j},1|{\Bbb E}%
_{k}\right] {\normalsize \varphi }=2f_{jkh}^{\left( -\right) }1|{\Bbb E}_{h}~%
{\normalsize \varphi \quad ,}
\end{equation}
where $f_{jkh}^{\left( \pm \right) }\left( \varphi \right) $ are the left
and right parallelizable torsion. One can check that our ${\Bbb E}_{i}$
defines what Cartan calls pure spinors \cite{r22}, 
\begin{equation}
\varphi ^{t}{\Bbb E}_{i}\varphi =0
\end{equation}
thus 
\begin{equation}
f_{ijk}^{\left( +\right) }\left( \varphi \right) =\frac{\varphi ^{t}\left( -%
{\Bbb E}_{k}{\Bbb E}_{i}{\Bbb E}_{j}\right) \varphi }{r^{2}}.
\end{equation}
and 
\begin{equation}
f_{ijk}^{\left( -\right) }\left( \varphi \right) =\frac{\varphi ^{t}\left(
-1|{\Bbb E}_{k}\;\;1|{\Bbb E}_{i}\;\;1|{\Bbb E}_{j}\right) \varphi }{r^{2}}.
\end{equation}
where 
\begin{equation}
\varphi ^{t}\varphi =r^{2}.
\end{equation}
There is another interesting and very important property to note 
\begin{equation}
\varphi ^{t}\left[ {\Bbb E}_{i},1|{\Bbb E}_{j}\right] \varphi =0  \label{mmm}
\end{equation}
which may be the generalization of the standard Lie algebra relation, left
and right action commute everywhere over the group manifold.

We close the algebra pointwisely using structure functions $f_{ijk}(\varphi
) $ instead of structure constants $f_{ijk}$ where $\varphi $ may be
considered as a coordinate system for an internal $S^{7}$ manifold not the
space-time $x$ and they don't mix 
\begin{equation}
\frac{\partial x}{\partial \varphi }=\frac{\partial \varphi }{\partial x}=0.
\end{equation}
Apart from the commutation of left and right actions, there are some other
useful identities satisfied by our $\left( {\Bbb E}_{j},1|{\Bbb E}%
_{j}\right) $ quaternionic or octonionic structures, they are 
\begin{eqnarray}
({\Bbb E}_{k})_{\mu \nu }({\Bbb E}_{j})_{\lambda \nu }+({\Bbb E}_{j})_{\mu
\nu }({\Bbb E}_{k})_{\lambda \nu } &=&2\delta _{kj}\delta _{\mu \lambda }, 
\nonumber \\
({\Bbb E}_{k})_{\mu \nu }({\Bbb E}_{j})_{\mu \lambda }+({\Bbb E}_{j})_{\mu
\nu }({\Bbb E}_{k})_{\mu \lambda } &=&2\delta _{kj}\delta _{\nu \lambda }, 
\nonumber \\
({\Bbb E}_{k})_{\mu \nu }({\Bbb E}_{k})_{\lambda \zeta }+({\Bbb E}%
_{k})_{\lambda \nu }({\Bbb E}_{k})_{\mu \zeta } &=&2\delta _{\mu \lambda
}\delta _{\nu \zeta },  \label{trl}
\end{eqnarray}
and the same holds equally well for $(1|{\Bbb E}_{j})$, as had been noticed
by Evans \cite{r4}, they are direct consequences of the ring division
triality.

Now, we have all the needed ingredients to construct our $\ $real universal $%
\left( \Gamma _{M}\right) _{ab}$ matrices with spinorial lower indices $a,b$
of range the double of the $\mu $. For Minkowskian metric of signature $\eta
\equiv \left( -,+,\ldots ,+\right) $, in $d=3,4,6$ and $10$, $a,b=0..2\mu +1$%
, for simplicity, we use symmetric $\Gamma _{M}$, 
\begin{equation}
\begin{array}{lll}
_{\left( \Gamma _{j}\right) _{ab}=\left( 
\begin{array}{cc}
0 & {\Bbb E}_{j} \\ 
-{\Bbb E}_{j} & 0
\end{array}
\right) } &  & _{\left( 1|\Gamma _{j}\right) _{ab}=\left( 
\begin{array}{cc}
0 & 1|{\Bbb E}_{\mu } \\ 
-1|{\Bbb E}_{j} & 0
\end{array}
\right) ,} \\ 
_{\left( \Gamma _{0}\right) _{ab}=\left( 
\begin{array}{cc}
-{\bf 1}_{\mu } & 0 \\ 
0 & -{\bf 1}_{\mu }
\end{array}
\right) ;} & _{\left( \Gamma _{d-2}\right) =\left( 
\begin{array}{cc}
0 & {\bf 1}_{\mu } \\ 
{\bf 1}_{\mu } & 0
\end{array}
\right) } & _{\left( \Gamma _{d-1}\right) _{ab}=\left( 
\begin{array}{cc}
{\bf 1}_{\mu } & 0 \\ 
0 & -{\bf 1}_{\mu }
\end{array}
\right) ,}
\end{array}
\label{gmm}
\end{equation}
The corresponding higher indices $\left( \tilde{\Gamma}\right) ^{ab}$'s are 
\begin{equation}
\left( \tilde{\Gamma}_{0}\right) ^{ab}=-\left( \Gamma _{0}\right)
_{ab}\qquad and\;\;\quad \quad \left( \tilde{\Gamma}\right) ^{ab}=\left(
\Gamma \right) _{ab}\quad .
\end{equation}
As a result, we find 
\[
\Gamma ^{M}\tilde{\Gamma}^{N}+\Gamma ^{N}\tilde{\Gamma}^{M}=1|\Gamma ^{M}1|%
\tilde{\Gamma}^{N}+1|\Gamma ^{N}1|\tilde{\Gamma}^{M}=2\eta ^{MN}
\]
or in terms of components 
\begin{equation}
_{\left( \Gamma ^{M}\right) _{ab}\left( \tilde{\Gamma}^{N}\right)
^{bc}+\left( \Gamma ^{N}\right) _{ab}\left( \tilde{\Gamma}^{M}\right)
^{bc}=\left( 1|\Gamma ^{M}\right) _{ab}\left( 1|\tilde{\Gamma}^{N}\right)
^{bc}+\left( 1|\Gamma ^{N}\right) _{ab}\left( 1|\tilde{\Gamma}^{M}\right)
^{bc}=2\eta ^{MN}\delta _{a}^{c}.}
\end{equation}
Our $\Gamma $'s satisfy the very important identity \cite{r4} 
\begin{equation}
\Gamma _{Ma(b}\Gamma _{cd)}^{M}=1|\Gamma _{Ma(b}1|\Gamma _{cd)}^{M}=0.
\label{ident}
\end{equation}

\section{The SSYM's Auxliary Fields}

Using Evans ansatz \cite{r4}, SSYM are composed of: Gauge fields $A_{M}$,
spinors $\Psi ^{a}$, $j\left( =1..d-3\right) $ algebraic auxiliary fields $%
K^{j}$. The gauge group indices will be suppressed in the following. The
Lagrangian density is 
\begin{equation}
{\cal L}=-\frac{1}{4}F_{MN}F^{MN}+\frac{i}{2}\Psi ^{t}\Gamma ^{M}\nabla
_{M}\Psi +\frac{1}{2}\delta _{ij}K^{i}K^{j},  \label{ddd}
\end{equation}
where $\nabla _{M}\equiv \partial _{M}+A_{M};$ $F_{MN}\equiv \left[ \nabla
_{M},\nabla _{N}\right] $ and the $\Gamma $ are given in (\ref{gmm}). The
Lagrangian is invariant up to a total derivative iff (\ref{ident}) holds.
Our supersymmetry transformations are\footnote{%
Contrary to \cite{r6}, we set $\Lambda _{j}=\tilde{\Lambda}^{j}$ from the
strart.} 
\begin{eqnarray}
\delta _{\eta }A_{M} &=&i\eta \Gamma _{M}\Psi ,  \nonumber \\
\delta _{\eta }\Psi ^{\alpha } &=&\frac{1}{2}F_{MN}\left( \Gamma _{MN}~\eta
\right) ^{\alpha }+K^{j}\left( \Lambda _{j}\right) _{\beta }^{\alpha }\eta
^{\beta },  \nonumber \\
\delta _{\eta }K_{j} &=&i\left( \Gamma ^{M}\nabla _{M}\Psi \right) _{\alpha
}\left( \Lambda _{j}\right) _{\beta }^{\alpha }\eta ^{\beta },  \nonumber \\
&&  \label{dddd}
\end{eqnarray}
where $\Lambda _{P}$ are some real matrices and Lorentz transformations are
generated by $\Gamma _{MN}\equiv \tilde{\Gamma}_{[M}\Gamma _{N]}$. Imposing
the closure of the supersymmetry infinitesimal transformations 
\begin{equation}
\left[ \delta _{\epsilon },\delta _{\eta }\right] =2i\epsilon ^{t}\Gamma
^{M}\eta \partial _{M}\;.  \label{alg}
\end{equation}
The closure on $A_{M}$ yields 
\begin{equation}
\Gamma _{M}\Lambda _{j}+\left( \Lambda _{j}\right) ^{t}\Gamma _{M}=0.
\label{c1}
\end{equation}
In addition to this condition the closure on $K^{j}$ also requires 
\begin{equation}
\Lambda _{j}\Lambda _{h}+\Lambda _{h}\Lambda _{j}=-2\delta _{jh}.  \label{c2}
\end{equation}
While closure on the fermionic field $\Psi ^{\alpha }$ holds iff 
\[
\left( \Gamma ^{M}\right) _{\alpha \beta }\left( \tilde{\Gamma}_{M}\right)
^{\gamma \delta }=2\delta _{(\alpha }^{\gamma }\delta _{\beta )}^{\delta
}+2\left( \Lambda _{j}\right) _{(\alpha }^{\gamma }\left( \Lambda
_{j}\right) _{\beta )}^{\delta }. 
\]
Now, we continue in a different way to Evans. To construct $\Lambda _{j}$,
we first impose the additional condition 
\begin{equation}
\left( \Lambda \right) ^{t}=-\left( \Lambda \right) ,  \label{add}
\end{equation}
we notice from (\ref{c2}) that the $\Lambda _{j}$ form a real Clifford
algebra, and from (\ref{c1}) 
\begin{equation}
\Gamma _{M}\Lambda _{j}-\Lambda _{j}\Gamma _{M}=0.
\end{equation}
that they commute with our space-time $\Gamma _{M}$ Clifford algebra. The
solution of the auxiliary field problem for $d=3,4,6$ dimensions, using (\ref
{gmm}) is then simply 
\begin{equation}
\Lambda _{j}=\left( 
\begin{array}{cc}
1|{\Bbb E}_{j} & 0 \\ 
0 & 1|{\Bbb E}_{j}
\end{array}
\right) ,  \label{lam}
\end{equation}
because 
\begin{equation}
\left\{ 1|{\Bbb E}_{j},1|{\Bbb E}_{h}\right\} =-2\delta _{jh},
\end{equation}
and 
\begin{equation}
\left[ {\Bbb E}_{j},1|{\Bbb E}_{h}\right] =0.
\end{equation}
Of course this solution is not unique. For example, if someone had started
with $1|\Gamma _{M}$, he would have found $\Lambda _{j}=\left( 
\begin{array}{ll}
{\Bbb E}_{j} & 0 \\ 
0 & {\Bbb E}_{j}
\end{array}
\right) $.

Now, we can relax the conditions (\ref{gmm}) and (\ref{add}). In general,
one replaces left/right action used for the gamma matrices by right/left
action for the $\Lambda _{j}$ e.g. 
\begin{equation}
_{\left( \Gamma _{j}\right) _{ab}=\left( 
\begin{array}{cc}
0 & {\Bbb E}_{j}|{\Bbb E}_{j+1} \\ 
-{\Bbb E}_{j}|{\Bbb E}_{j+1} & 0
\end{array}
\right) \;\ \rightarrow \;\;\left( \Lambda _{j}\right) _{ab}=\left( 
\begin{array}{cc}
{\Bbb E}_{j+1}|{\Bbb E}_{j} & 0 \\ 
0 & {\Bbb E}_{j+1}|{\Bbb E}_{j}
\end{array}
\right) }
\end{equation}
One writes any $\Gamma $ and expand it in terms left/right action $\left( 
{\Bbb E}_{i,}1|{\Bbb E}_{j},{\Bbb E}_{m}|{\Bbb E}_{n}\right) $ then the $%
\Lambda $ will be given in terms of suitable $\left( 1|{\Bbb E}_{i,}{\Bbb E}%
_{j},{\Bbb E}_{n}|{\Bbb E}_{m}\right) $ taking into account that daigonal
elements should be replaced by non-diagonal one and interchanging left/right
actions simultaneously.

\section{The Ten Dimensions Case}

For $d=10$, working with octonions the situation is different. We know that
octonionic left and right action commutes only when applied to $\varphi $, 
\begin{equation}
\varphi ^{t}\left[ {\Bbb E}_{j},1|{\Bbb E}_{h}\right] \varphi =0,
\end{equation}
and $\varphi $ is just an 8 dimensional column matrix. Up to now, we have
not restricted $\varphi $ by any other conditions. With two different $%
\varphi $, $\left( \varphi ^{\left( 1\right) },\varphi ^{\left( 2\right)
}\right) $, we impose now the conditions that $\varphi ^{\left( i\right) }$
be fermionic fields. We express our 16 dimensional Grassmanian variables $%
\epsilon ,\eta $ of eqn.(\ref{alg}) in terms of $\varphi $, 
\begin{equation}
\begin{array}{lll}
& \epsilon =\eta ^{t} &  \\ 
& \;\Downarrow &  \\ 
\epsilon =\left( 
\begin{array}{ll}
\varphi ^{\left( 1\right) } & \varphi ^{\left( 2\right) }
\end{array}
\right) ; &  & \eta =\left( 
\begin{array}{l}
\varphi ^{\left( 1\right) } \\ 
\varphi ^{\left( 2\right) }
\end{array}
\right)
\end{array}
\label{cond1}
\end{equation}
We now rederive (\ref{alg}) for the octonions. The closure conditions of our
algebra, {\em without omitting the Grassmanian variables} are 
\begin{eqnarray}
\eta ^{t}\left( \Gamma _{M}\Lambda _{j}-\Lambda _{j}\Gamma _{M}\right) \eta
&=&0,  \nonumber \\
\eta ^{t}\left( \Lambda _{j}\Lambda _{h}+\Lambda _{h}\Lambda _{j}\right)
\eta &=&\eta ^{t}\left( -2\delta _{jh}\right) \eta ,  \nonumber \\
\eta ^{t}\left( \left( \Gamma ^{M}\right) _{\alpha \beta }\left( \tilde{%
\Gamma}_{M}\right) ^{\gamma \delta }\right) \eta &=&\eta ^{t}\left( 2\delta
_{(\alpha }^{\gamma }\delta _{\beta )}^{\delta }+2\left( \Lambda _{j}\right)
_{(\alpha }^{\gamma }\left( \Lambda _{j}\right) _{\beta )}^{\delta }\right)
\eta ,  \label{algoct}
\end{eqnarray}
which are satisfied for the octonionic representation 
\begin{equation}
\left( \Gamma _{j}\right) _{ab}=\left( 
\begin{array}{cc}
0 & {\Bbb E}_{j} \\ 
-{\Bbb E}_{j} & 0
\end{array}
\right) ,\quad \quad \Lambda _{j}=\left( 
\begin{array}{cc}
1|{\Bbb E}_{j} & 0 \\ 
0 & 1|{\Bbb E}_{j}
\end{array}
\right) .
\end{equation}
By interchanging left/right action, we have different solutions as in the
quaternionic case. In summary, while the fermionic fields couple to
left/right action through the gamma matrices, the auxiliary fields couple to
right/left action through the $\Lambda $. For the octonionic case {\em the
presence of the Grassmanian variables is essential.} Contrary to the
standard supersymmetry transformation, our Grassman variables are the same ($%
\epsilon =\eta ^{t}$), which is identical to the result obtained by
Berkovits in \cite{r7}. \ According to Evans \cite{r8}, the attractive
feature of this scheme is that the Lagrangian (\ref{ddd}) and the
transformation (\ref{dddd}) are manifestly invariant under the generalized
Lorentz group $SO\left( 1,9\right) $. In our formulation, we can show some
additional characteristic. In some cases, the \ (\ref{cond1}) condition may
be relaxed, for equal $j$ or $h$ (no summation) 
\begin{equation}
\left. 
\begin{array}{c}
\varphi ^{t}\;{\Bbb E}_{j}\;\left[ {\Bbb E}_{j},1|{\Bbb E}_{h}\right] \varphi
\\ 
\varphi ^{t}\;\;1|{\Bbb E}_{i}\;\;\left[ {\Bbb E}_{j},1|{\Bbb E}_{h}\right]
\varphi \\ 
\varphi ^{t}\;E_{h}\;\left[ {\Bbb E}_{j},1|{\Bbb E}_{h}\right] \varphi \\ 
\varphi ^{t}\;\;1|E_{h}\;\;\left[ {\Bbb E}_{j},1|{\Bbb E}_{h}\right] \varphi
\end{array}
\right\} =0.
\end{equation}
i.e. relating $\epsilon $ and $\eta $ by an $S^{7}$ is also allowed.

Now, Let us show what will happen to $spin\left( 1,9\right) $ when we
transform it to $soft\;spin\left( 1,9\right) $ 
\begin{eqnarray}
soft\;spin\left( 1,9\right) &\sim &\left[ \Gamma _{i},\Gamma _{j}\right] \eta
\nonumber \\
&=&\left[ \left( 
\begin{array}{cc}
0 & {\Bbb E}_{i} \\ 
-{\Bbb E}_{i} & 0
\end{array}
\right) ,\left( 
\begin{array}{cc}
0 & {\Bbb E}_{j} \\ 
-{\Bbb E}_{j} & 0
\end{array}
\right) \right] \left( 
\begin{array}{l}
\varphi ^{\left( 1\right) } \\ 
\varphi ^{\left( 2\right) }
\end{array}
\right)  \nonumber \\
&=&-\left( 
\begin{array}{cc}
0 & \left[ {\Bbb E}_{i},{\Bbb E}_{j}\right] \\ 
\left[ {\Bbb E}_{i},{\Bbb E}_{j}\right] & 0
\end{array}
\right) \left( 
\begin{array}{l}
\varphi ^{\left( 1\right) } \\ 
\varphi ^{\left( 2\right) }
\end{array}
\right)  \nonumber \\
&=&-\left( 
\begin{array}{ll}
0 & f_{ijk}^{\left( +\right) }\left( \varphi ^{\left( 2\right) }\right) 
{\Bbb E}_{k} \\ 
f_{ijk}^{\left( +\right) }\left( \varphi ^{\left( 1\right) }\right) {\Bbb E}%
_{k} & 0
\end{array}
\right) \left( 
\begin{array}{l}
\varphi ^{\left( 1\right) } \\ 
\varphi ^{\left( 2\right) }
\end{array}
\right) .  \nonumber \\
&&
\end{eqnarray}

\section{Some Superspace Hints}

Lastly, let us make some comments about a possible superspace. It seems that
the best way to find the $d=6,10$ superspace for SSYM is by defining some
quaternionic and octonionic Grassmann variables that decompose the
corresponding spinors into an $SL\left( 2,H\right) $ and an $SL\left(
2,soft\;S^{7}\right) $ respectively 
\begin{equation}
\left\{ \theta _{\alpha },\theta _{\beta }\right\} =\left\{ \bar{\theta}_{%
\dot{\alpha}},\bar{\theta}_{\dot{\beta}}\right\} =\left\{ \theta _{\alpha },%
\bar{\theta}_{\dot{\beta}}\right\} =0,
\end{equation}
where $\alpha =1,2$ over quaternions or octonions. We know that the
supersymmetry generators $Q_{\alpha }$ are derived from right multiplication 
\begin{eqnarray}
{Q}_{\alpha } &=&\left( \partial _{\alpha }-1|\Gamma _{\alpha \dot{\beta}%
}^{\mu }\bar{\theta}^{\dot{\beta}}P_{\mu }\right)  \\
{Q}^{\alpha } &=&\left( -\partial ^{{\alpha }}+\bar{\theta}_{\dot{\beta}}1|%
\tilde{\Gamma}^{\mu \dot{\beta}\alpha }P_{\mu }\right) 
\end{eqnarray}
also 
\begin{equation}
{\bar{Q}}^{\dot{\alpha}}=\left( \partial ^{\dot{\alpha}}-1|\tilde{\Gamma}%
^{\mu \dot{\alpha}\alpha }\theta _{\alpha }P_{\mu }\right) 
\end{equation}
\begin{equation}
{\bar{Q}}_{\dot{\alpha}}=\left( -\partial _{\dot{\alpha}}+\theta ^{\alpha
}1|\Gamma _{\alpha \dot{\alpha}}P_{\mu }\right) 
\end{equation}
whereas the covariant derivative $D_{\alpha }$ are obtained by left action 
\begin{eqnarray}
{D}_{\alpha } &=&\left( \partial _{\alpha }+\Gamma _{\alpha \dot{\beta}%
}^{\mu }\bar{\theta}^{\dot{\beta}}P_{\mu }\right)  \\
{D}^{\alpha } &=&\left( -\partial ^{{\alpha }}-\bar{\theta}_{\dot{\beta}}%
\tilde{\Gamma}^{\mu \dot{\beta}\alpha }P_{\mu }\right) 
\end{eqnarray}
also 
\begin{equation}
{\bar{D}}^{\dot{\alpha}}=\left( \partial ^{\dot{\alpha}}+\tilde{\Gamma}^{\mu 
\dot{\alpha}\alpha }\theta _{\alpha }P_{\mu }\right) 
\end{equation}
\begin{equation}
{\bar{D}}_{\dot{\alpha}}=\left( -\partial _{\dot{\alpha}}-\theta ^{\alpha
}\Gamma _{\alpha \dot{\alpha}}P_{\mu }\right) 
\end{equation}
Leading to a result acceptable but different from the standard $N=1$, $d=4$
superspace, 
\begin{eqnarray}
\{Q_{\alpha },\bar{Q}_{\dot{\alpha}}\} &=&-2\left( 1|\Gamma _{\alpha {\dot{%
\alpha}}}^{\mu }\right) P_{\mu }\;,  \nonumber \\
\{Q_{\alpha },Q_{\beta }\} &=&\{\bar{Q}_{\dot{\alpha}},\bar{Q}_{\dot{\beta}%
}\}\;=\;0\;,  \nonumber
\end{eqnarray}
\begin{eqnarray}
\{D_{\alpha },\bar{D}_{\dot{\alpha}}\} &=&2\Gamma _{\alpha {\dot{\alpha}}%
}^{\mu }P_{\mu }\;,  \nonumber \\
\{D_{\alpha },D_{\beta }\} &=&\{\bar{D}_{\dot{\alpha}},\bar{D}_{\dot{\beta}%
}\}\;=\;0\;,  \nonumber
\end{eqnarray}
and iff left and right action commute, we restore 
\begin{eqnarray}
\{Q_{\alpha },\bar{D}_{\dot{\alpha}}\} &=&\{D_{\alpha },\bar{Q}_{\dot{\alpha}%
}\}=0\;,  \nonumber \\
\{Q_{\alpha },D_{\beta }\} &=&\{\bar{D}_{\dot{\alpha}},\bar{Q}_{\dot{\beta}%
}\}\;=\;0\;.  \nonumber
\end{eqnarray}
On the other hand for octonions we would have the weaker conditions, 
\begin{eqnarray}
\left( 
\begin{array}{ll}
\varphi ^{\left( 1\right) } & \varphi ^{\left( 2\right) }
\end{array}
\right) \{Q_{\alpha },\bar{D}_{\dot{\alpha}}\}\left( 
\begin{array}{l}
\varphi ^{\left( 1\right) } \\ 
\varphi ^{\left( 2\right) }
\end{array}
\right)  &=&\left( 
\begin{array}{ll}
\varphi ^{\left( 1\right) } & \varphi ^{\left( 2\right) }
\end{array}
\right) \{D_{\alpha },\bar{Q}_{\dot{\alpha}}\}\left( 
\begin{array}{l}
\varphi ^{\left( 1\right) } \\ 
\varphi ^{\left( 2\right) }
\end{array}
\right) =0\;,  \nonumber \\
\left( 
\begin{array}{ll}
\varphi ^{\left( 1\right) } & \varphi ^{\left( 2\right) }
\end{array}
\right) \{Q_{\alpha },D_{\beta }\}\left( 
\begin{array}{l}
\varphi ^{\left( 1\right) } \\ 
\varphi ^{\left( 2\right) }
\end{array}
\right)  &=&\left( 
\begin{array}{ll}
\varphi ^{\left( 1\right) } & \varphi ^{\left( 2\right) }
\end{array}
\right) \{\bar{D}_{\dot{\alpha}},\bar{Q}_{\dot{\beta}}\}\left( 
\begin{array}{l}
\varphi ^{\left( 1\right) } \\ 
\varphi ^{\left( 2\right) }
\end{array}
\right) =\;0\;.  \nonumber
\end{eqnarray}
The commutation of left and right actions is not just needed for
associativity but for the invariance under supersymmetry transformation 
\begin{equation}
\delta _{\xi }\equiv \xi Q+\bar{\xi}\bar{Q}
\end{equation}
because only the associativity ensures 
\begin{equation}
\left( 
\begin{array}{ll}
\varphi ^{\left( 1\right) } & \varphi ^{\left( 2\right) }
\end{array}
\right) \left[ \delta _{\xi },D_{\alpha }\right] \left( 
\begin{array}{l}
\varphi _{1} \\ 
\varphi _{2}
\end{array}
\right) =\left( 
\begin{array}{ll}
\varphi ^{\left( 1\right) } & \varphi ^{\left( 2\right) }
\end{array}
\right) \left[ \delta _{\xi },\bar{D}_{\dot{\alpha}}\right] \left( 
\begin{array}{l}
\varphi _{1} \\ 
\varphi _{2}
\end{array}
\right) =0,
\end{equation}
since $\delta _{\xi }$ is left action and $D_{\alpha }$ is right action
which is a very important relation in the standard $N=1$ superspace for the
invariance of the Lagrangian under supersymmetry transformation. We hope to
return to this point in a future work.\newline
\newline
\newline
\newline
\newline
I am grateful to C. Imbimbo, P. Rotelli and A. Van Proeyen for useful
comments.


\newpage 

\begin{thebibliography}{99}
\bibitem{r0}  L.~Brink,~J.~Scherk~and~J.~H.~Schwarz, Nucl.~Phys.~{\bf 121B}
(1977) 77.

\bibitem{r1}  J.~H.~Schwarz, ``Introduction To Supersymmetry'', Presented at
28th Scottish Universities Summer School in Physics, Edinburgh, Scotland,
Jul 28 - Aug 17, 1985.

\bibitem{r2}  T.~Kugo and P.~Townsend, Nucl. Phys. {\bf B221} (1983) 357.

\bibitem{r3}  K.~W.~Chung~and~A.~Sudbery, Phys.~Lett. {\bf B198} (1987) 161.

\bibitem{r4}  J.~M.~Evans, Nucl.~Phys. {\bf B298} (1988) 92.

\bibitem{r6}  J.~M.~Evans, Nucl.~Phys. {\bf B310} (1988) 44 .

\bibitem{r7}  N.~Berkovits, Phys. Lett. {\bf B318} (1993) 104.

\bibitem{r8}  J.~M.~Evans, Phys. Lett. {\bf B334} (1994) 105. J.~M.~Evans,
STRINGS AND SYMMETRIES, Edited by Gulen Aktas et al, Springer-Verlag
(Lecture Notes in Physics, 447) 1995.

\bibitem{r9}  I.~Bengtsson, M.~Cederwall, Nucl.~Phys. {\bf B302}, (1988) 81.

\bibitem{r10}  H.~Tachibana, K.~Imaeda, Nuovo Cimento {\bf B104}, (1989) 91.
D.~B.~Fairlie and C.A.~Manogue, Phys.~Rev. {\bf D36} (1987) 475,
C.A.~Manogue, A.~Sudbery, Phys.~Rev. {\bf D40} (1989) 4073. I.~Oda,
T.~Kimura, A.~Nakamura, Prog.~Theor.~Phys.~{\bf 80} (1988) 367.
M.~Cederwall, Phys. Lett. {\bf B210} (1988) 169, E.~Corrigan,
T.J.~Hollowood, Commun.~Math.~Phys.~{\bf 122} (1989) 393, E.~Corrigan,
T.J.~Hollowood, Phys. Lett. {\bf B203}, (1988) 47, R.~Foot, G.C.~Joshi,
Int.~J.~Theor.~Phys.~{\bf 28} (1989) 1449, R.~Foot, G.C.~Joshi, Phys. Lett. 
{\bf B199}, (1987) 203.

\bibitem{r11}  {M.P.~Blencowe and M.J.~Duff, Nucl.~Phys.~{\bf B310} (1988)
387.}

\bibitem{r12}  B.~E.~Nilsson, ``Off-Shell Fields For The Ten-Dimensional
Supersymmetric Yang-Mills Theory,'' GOTEBORG-81-6.

\bibitem{r13}  B.~E.~Nilsson, Class.~Quant.~Grav.~{\bf 3} (1986) L41.

\bibitem{r14}  K. Abdel-Khalek, Int. J. Mod. Phys. A13 (1998) 569.

\bibitem{r15}  K. Abdel-Khalek, ``Ring Divsion Algebras, Self--Duality and
Supersymmetry'', Ph.D. Lecce University, Feb. 2000.

\bibitem{r16}  P.~Rotelli, Mod.~Phys.~Lett. A., {\bf 4} (1989) 933.

\bibitem{r17}  G.~'t~Hooft, Phys.~Rev. {\bf 14D} (1976) 3432.

\bibitem{r18}  E. Cartan and J. A. Schouten, Proc. Kon. Akad. Wet.
Amesterdam 29 (1926) 803, 933.

E. Cartan, J. Math. Pures et Appl. 6 (1927) 1.

\bibitem{r18a}  M.~Sohnius, Z. Phys. {\bf C18} (1983) 229.

\bibitem{r19}  F.~Englert, A.~Servin, W.~Troost, A.~Van~Proeyen and
Ph.~Spindel, J.~Math.~Phys. {\bf 29} (1988) 281.

\bibitem{r20}  K. Abdel-Khalek, math-ph/0002024.

\bibitem{r22}  P. Budinich, Commun. Math. Phys. 107 (1986) 455.
\end{thebibliography}
\end{document}